\documentclass[manuscript]{aastex61}
\usepackage{graphicx}
\usepackage{subfigure}
\usepackage{longtable}
\usepackage{multirow}
\usepackage{xcolor}

\shorttitle{}
\shortauthors{}

\begin{document}

\title{Massive White Dwarfs in the Galactic Center: A Chandra X-ray Spectroscopy of Cataclysmic Variables}

\author{Xiao-jie Xu}
\affil{School of Astronomy and Space Science, Key Laboratory of Modern Astronomy and Astrophysics, Nanjing University, Nanjing, P. R. China 210046}
\email{xuxj@nju.edu.cn}
\author{Zhiyuan Li}
\affil{School of Astronomy and Space Science, Key Laboratory of Modern Astronomy and Astrophysics, Nanjing University, Nanjing, P. R. China 210046}
\author{Zhenlin Zhu}
\email{}
\affil{School of Astronomy and Space Science, Key Laboratory of Modern Astronomy and Astrophysics, Nanjing University, Nanjing, P. R. China 210046}
\author{Zhongqun Cheng}
\email{}
\affil{Department of Astronomy, Xiamen University, Xiamen, Fujian, Department of Astronomy, Xiamen University, Xiamen, Fujian, P. R. China 361005}
\author{Xiang-dong Li}
\affil{School of Astronomy and Space Science, Key Laboratory of Modern Astronomy and Astrophysics, Nanjing University, Nanjing, P. R. China 210046}
\author{Zhuo-li Yu}
\email{}
\affil{School of Astronomy and Space Science, Key Laboratory of Modern Astronomy and Astrophysics, Nanjing University, Nanjing, P. R. China 210046}

\begin{abstract}
Previous X-ray observations toward the Nuclear Star Cluster (NSC) at the Galactic center have discovered thousands of point sources, most of which were believed to be cataclysmic variables (CVs), i.e., a white dwarf (WD) accreting from a low-mass companion. However, the population properties of these CVs remain unclear, which otherwise contain important information about the evolutionary history of the NSC. 
In this work we utilize ultradeep archival \textit{Chandra} observations to study the spectral properties of the NSC CVs, in close comparison with those in the Solar vicinity. 
We find that the NSC CVs have strong Fe XXV and Fe XXVI lines (both of which show equivalent widths $\sim200-300$ eV), indicating metal-rich companions. Moreover, their Fe XXVI to Fe XXV line flux ratio is used to diagnose the characteristic white dwarf mass ($M_{\rm WD}$) of NSC CVs. The results show that the CVs with $L_{\rm 2-10 keV}>6\times10^{31}$ erg s$^{-1}$ have a mean $M_{\rm WD}$ of $\sim0.6/1.0\,M_{\odot}$ if they are magnetic/non-magnetic CVs; while those with $L_{\rm 2-10 keV}$ between  $1-6\times10^{31}$ erg s$^{-1}$ have a mean $M_{\rm WD}$ of $\sim0.8/1.2\,M_{\odot}$ if they are magnetic/non-magnetic CVs. All these \textit{Chandra}-detected CVs collectively contribute $\sim$30-50\% of the unresolved 20-40 keV X-ray emission from the NSC. The CV population with massive (i.e., $M_{\rm WD}\sim1.2M_{\odot}$) WDs have not been observed in the Solar vicinity or the Galactic bulge, and they might have been formed via dynamical encounters in the NSC.

\end{abstract}

\keywords{binaries: close --- X-rays: binaries ---Galaxy: center --- stars: kinematics and dynamics}

\section{Introduction}
Consisting of a vast number ($\sim$$10^{7}$) of predominantly old stars densely concentrated in the innermost few parsecs of our Galaxy, the Nuclear Star Cluster (NSC) provides an important laboratory for the understanding of fundamental astrophysics (see \citealt{genzel2010} for a recent review). In particular, how individual stars and binaries would evolve under the influence of their mutual dynamics, which is persistently regulated by the gravity of a central super-massive black hole (SMBH).

In this work, we concentrate on the population properties of cataclysmic variables (CVs), in which a white dwarf (WD) accretes matter from its main-sequence (MS) or sub-giant companion star and emits X-rays. CVs are good targets to study the stellar evolution theory, and they are closely related to more interesting astrophysical objects, like the progenitors of type Ia supernovae, which are believed to be binaries containing one or two WDs \citep{wang2012}, and close double WD binaries, which are main targets for future gravitational wave detectors like TianQin \citep{luo2016}. 

In the past two decades, X-ray observations provide an important approach to study the CV in the NSC. X-ray photons from CVs with energy above $\sim2$ keV can penetrate the foreground absorbing gas and provide information on the binary population and therefore the changes due to dynamical effects in the NSC. For example,  the specific X-ray luminosity function of point sources (normalized by total stellar mass) in the NSC region shows the enhanced abundance of X-ray sources above $\sim10^{31}$ erg s$^{-1}$  \citep[e.g.,][]{muno2009,zhu2018}, compared to those in the field. The combined X-ray spectra of point sources in the GC region resemble those of CVs in Solar vicinity, suggesting the origin of these sources should be CVs, just like the Galactic Bulge/Ridge X-ray emission(GB/RXE) \citep{muno2003,muno2009,zhu2018,rev09,sazonov2006}. Moreover, recent \textit{NuSTAR} observations have revealed an extended 20-40 keV hard X-ray emission in NSC field, which was named central hard X-ray emission  \citep[CHXE,][]{perez2015,hailey2016} in comparison to the well known Galactic center X-ray emission (GCXE). The broad band 2-40 keV spectrum of CHXE is consistent with spectra of magnetic CVs (mCVs, including polars and intermediate polars, aka, IPs) in the Solar vicinity  \citep{hailey2016}. The average shock temperature ($T_{\rm max}$) and mass of the WDs ($M_{\rm WD}$) were constrained to be $\sim$40 keV and $\sim0.9M_{\odot}$, respectively. Based on the luminosity function of point sources detected by previous \textit{Chandra} observations in the same region, \citet{hailey2016} further proposed that several thousands of mCVs, more specifically, IPs with 2-10 keV luminosity down to $\sim5\times10^{31}$ erg s$^{-1}$ would explain the CHXE.

Using a total of 4.4 Ms \textit{Chandra} observations, \citet{zhu2018} pushed the 2-10 keV detection limit to $\sim10^{31}$ erg s$^{-1}$ (assuming a distance of 8 kpc) in the NSC. The point sources within $250\arcsec$ and $L_{\rm X}$ below $6.0\times10^{31}$ erg s$^{-1}$ show strong H-like and He-like Fe emission lines (centered at $\sim6.97$ keV and $\sim6.7$ keV, respectively), which is consistent with typical IPs. However, the luminosities of these \textit{Chandra} detected sources are below the typical luminosity of magnetic CVs in the Solar vicinity ($\gtrsim10^{32}$ erg s$^{-1}$). 
As a result, their exact origin and the mean WD mass remain to be explored.

The flux ratio of Fe XXVI to Fe XXV emission lines ($I_{\rm 7.0}/I_{\rm 6.7}$) can be taken as a sensitive diagnostic for $T_{\rm max}$ and $M_{\rm WD}$ for CVs \citep[][see also Section 3 for details]{xu2016,yu2018,xu2019}. This is because a more massive WD would have a higher $T_{\rm max}$, thus more hydrogen-like Fe ions and a higher $I_{\rm 7.0}/I_{\rm 6.7}$. In this work, we use the $I_{\rm 7.0}/I_{\rm 6.7}$--$T_{\rm max}$--$M_{\rm WD}$ relations 
examined by IPs and non-mCVs in the Solar vicinity by \citet{xu2019} to diagnose the CV populations
in the NSC. We describe our data and method in Section 2. We compare the Fe line properties of the GCXE and the CHXE sources to those of CVs in Solar vicinity in Section 3, we explore the CV population in the NSC in section 4 and summarize in section 5.
Throughout this work, we quote errors at 90\% confidence level, unless otherwise stated.

\section{X-ray Data \& Analysis}

Our data reduction procedure is as described in \citet{zhu2018}, which presents a catalog of more than 3500 X-ray sources located in the inner 20 pc region of the GC, based on ultra-deep {\it Chandra} observations taken with the Advanced CCD Imaging Spectrometer (ACIS).
Since we are primarily interested in the spectral properties of CVs, other classes of X-ray sources in this catalog, e.g., X-ray transients (most likely LMXBs) and colliding wind massive binaries, as well as extended sources, have been excluded (see \citealp{zhu2018} for details of source identification). Any residual non-CV sources, in particular quiescent low mass X-ray binaries (qLMXBs) that are typically devoid of significant Fe lines, were estimated to be $\lesssim$ 5\% in number and should not significantly affect our results.
\begin{figure}[htbp]
\centering
\includegraphics[scale=0.4]{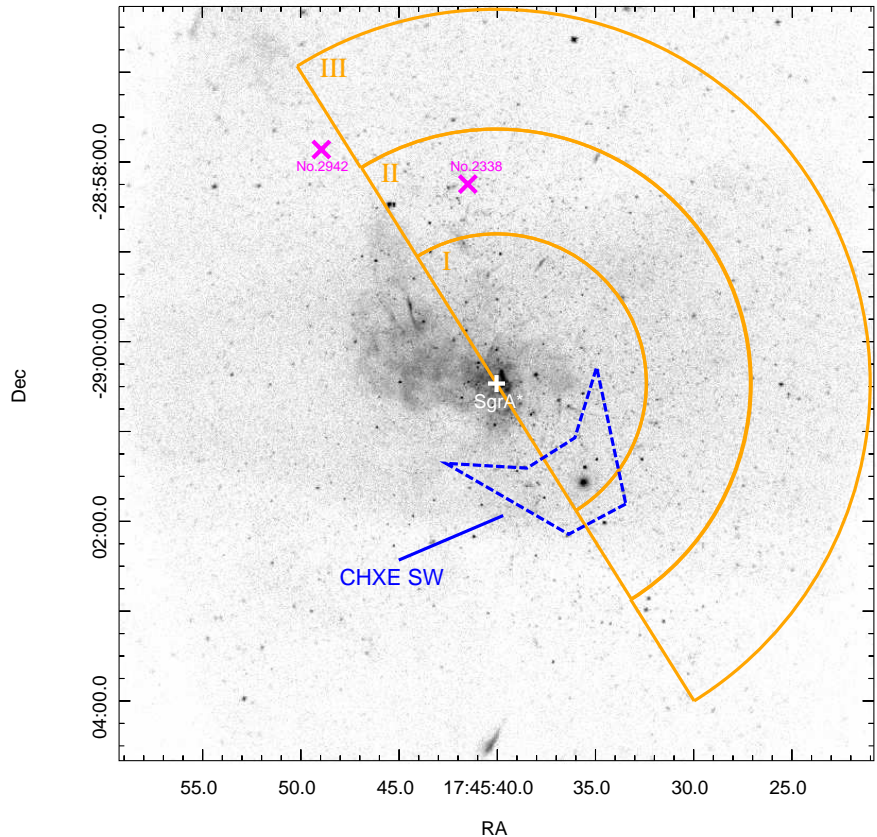}
   \caption{\textit{Chandra} 2--8 keV image of the $500\arcsec\times500\arcsec$ region centered at Sgr A*, from \citet{zhu2018}. The blue polygon denotes the CHXE-SW region defined by \citet{perez2015}, the pink crosses denote the positions of the two brightest sources from \citet{zhu2018}'s catalog, and the three orange half-annuli are regions I, II and III, in order of increasing radius.}
\label{fig:region}
\end{figure}

We focus on the GCXE region, defined here as the half-circle with a projected galacto-centric radius $R = 250\arcsec$ and Galactic latitude $b>0$, the latter criterion adopted to minimize the diffuse background (Figure~\ref{fig:region}). 
We then divide the point sources detected therein into two groups according to their luminosities:
GCXE-H (GCXE-L) consists of sources having 2--10 keV unabsorbed luminosity above (below) $6\times10^{31}{\rm~erg~s^{-1}}$, as measured in \citet{zhu2018}. 
Most IPs (non-mCVs) in the Solar vicinity are found above (below) this luminosity threshold \citep[e.g.,][]{xu2016}.  
The H and L sources are further divided into three sub-groups, according to their projected distances to Sgr A*: region I for $1\arcsec < R < 100\arcsec$, region II for $100\arcsec < R < 170\arcsec$, region III for $170\arcsec < R < 250\arcsec$. 
Finally, to compare with the \textit{NuSTAR} results, we select those {\it Chandra} sources falling within the CHXE-SW region as defined in \citet{perez2015}.
The spatial occupation of various sub-groups are illustrated in Figure~\ref{fig:region}.
For a given sub-group, we extract the cumulative spectra for each ObsID, using the CIAO tool {\it specextract}, and combine them with {\it combine\_spectra}. While both ACIS-I and ACIS-S data were utilized by \citet{zhu2018} for source detection, here we use only the 3\,Ms ACIS-S data to ensure an optimal spectral resolution for the Fe lines (including obs-IDs of 13850, 14392, 14394, 14393, 13856, 13857, 13854, 14413, 13855, 14414, 13847, 14427, 13848, 13849, 13846, 14438, 13845, 14460, 13844, 14461, 13853, 13841, 14465, 14466, 13842, 13839, 13840, 14432, 13838, 13852, 14439, 14462, 14463, 13851, 15568, 13843, 15570 and 14468. see \citealt{zhu2018} for detailed reasoning). 
The number of sources in each sub-group ranges from $99$ to $462$, sufficiently large to ensure that none of the cumulative spectra is dominated by just few sources.    

Following \citet{xu2016} and \citet{zhu2018}, the 3--8 keV continuum are then fitted with a phenomenological bremsstrahlung model. To account for the Fe lines, we make use of the 3-Gaussian model by \citet{xu2016} which was specifically constructed for this purpose. The parameters of this model include the centroid energies, widths, and (relative) intensities ($I_{6.4}/I_{6.7}$, $I_{6.7}$, and $I_{7.0}/I_{6.7}$) of the lines. This model automatically accounts for the correlations between the parameters of the Fe lines in error measurements, which can effectively reduce the error ranges of $I_{7.0}/I_{6.7}$ for further comparison. Both of the above components subject to an absorption column of order $10^{23}{\rm~cm^{-2}}$ (including foreground and intrinsic partial absorption).  

\begin{figure}[htbp]
  \centering
 \includegraphics[scale=0.5,angle=270]{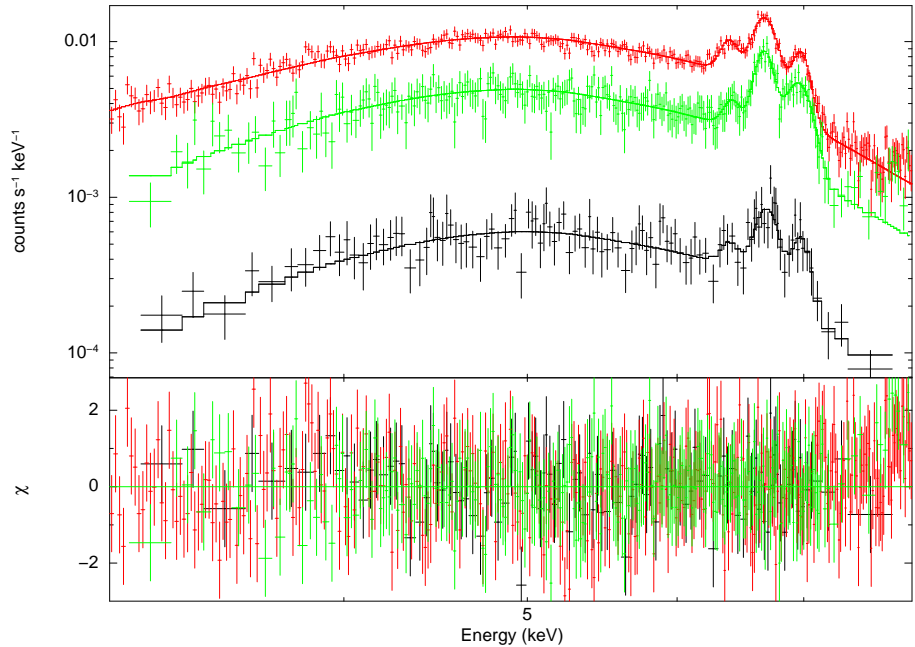}
  \caption{The cumulative \textit{Chandra} spectra of CHXE (black), GCXE-H (red) and GCXE-L (green) sources. The spectra are fitted by an absorbed bremsstrahlung + 3-Gaussian model to characterize the three Fe emission lines. The error bars are of 1-$\sigma$.}
  \label{fig:spec}
\end{figure}

The GCXE-H, GCXE-L and CHXE spectra are shown in the upper panel of Figure~\ref{fig:spec}.
Equivalent widths (EWs) and line flux ratios are derived from the spectral fit and presented in Table 1.  
It can be seen that the EW of Fe\,XXVI ($EW_{7.0}$), EW of Fe\,XXV ($EW_{6.7}$) and Fe\,XXVI to Fe\,XXV line flux ratio ($I_{7.0}/I_{6.7}$) are systematically higher in the L sources than in the H sources.
On the other hand, the $I_{\rm 7.0}/I_{\rm 6.7}$ values are consistent to within uncertainties among regions I, II and III; a similar conclusion can be drawn for the EWs ($EW_{7.0}$ of region III is marginally higher than the other two regions for the H sources). 
Therefore, we conclude that there is no significant radial gradient in the line ratio or EWs, and will not further distinguish these three sub-groups.

\begingroup
\begin{table}[htbp]
	\begin{center}
	\scriptsize
  \caption{Observed and derived properties of CVs in various environments}
  \centering
  \label{tbl:log}
\renewcommand{\arraystretch}{0.8} 
  \begin{tabular}{cccccccc}
    \hline
    Source & $EW_{6.4}$  & $EW_{6.7}$ & $EW_{7.0}$ & $I_{7.0/6.7}$  & $T_{\rm max}^{a}$ & $M_{\rm WD}^{b}$& $\chi^2$/dof\\
    &(eV) & (eV) &  (eV)&  & (keV)&($M_{\odot}$) &\\
    \hline
CHXE   & $46^{+42}_{-37}$ & $286^{+80}_{-70}$ & $190^{+70}_{-67}$  & $0.65\pm0.20$  & $23^{+10}_{-8}$ & $0.61^{+0.19}_{-0.14}$/$1.07^{+0.13}_{-0.16}$ &$0.72/95$\\
	    GCXE   & $61^{+5}_{-10}$ & $297\pm30$ & $206^{+10}_{-10}$  & $0.71\pm0.06$ & $26^{+2}_{-3}$  & $0.65^{+0.07}_{-0.02}$/$1.11^{+0.06}_{-0.03}$ &$0.79/268$\\
    GCXE-H  &$66\pm7$ & $269\pm15$ & $164\pm15$ & $0.60\pm0.05$ & $21^{+3}_{-3}$ & $0.57^{+0.05}_{-0.02}$/$1.04^{+0.04}_{-0.04}$ & $0.95/336$\\
    GCXE-L  & $41\pm17$ & $308\pm32$ & $310\pm36$ & $0.92\pm0.13$ & $38^{+11}_{-9}$ & $0.83^{+0.11}_{-0.11}$/$1.25^{+0.07}_{-0.08}$ &$0.72/250$\\
H$_{\rm I}$ & $77\pm26$ & $207\pm60$ & $120\pm30$  & $ 0.56_{-0.27}^{+0.23}$ & - & - & $0.91/273$\\
H$_{\rm II}$ & $73\pm28$ & $232\pm45$ & $110\pm40$   & $0.47\pm0.17 $ & - & - &$1.1/272$\\
H$_{\rm III}$ & $86\pm29$ & $221\pm24$ & $191\pm30$  & $0.76\pm0.19 $ & - & - &$0.93/261$\\
L$_{\rm I}$ & $<44$ & $580\pm220$ & $470\pm190$   & $1.0_{-0.4}^{+0.6} $ & - & -& $1.3/11$\\
L$_{\rm II}$ & $<42$ & $370\pm80$ & $269\pm87$   & $0.71_{-0.2}^{+0.3} $ & - & - &$0.78/116$\\
L$_{\rm III}$ & $77_{-34}^{+103}$ & $262\pm110$ & $300_{-140}^{+310}$  & $1.09\pm0.4 $ & - & -&$0.87/98$ \\
    \hline
	No.2338   & $102^{+45}_{-54}$ & $322^{+94}_{-96}$ & $123^{+81}_{-67}$  & $0.30^{+0.28}_{-0.17}$  & $11^{+8}_{-4}$ & $0.36^{+0.22}_{-0.11}$/$0.77^{+0.25}_{-0.30}$ & $1.1/76$\\
	No.2942   & $153^{+61}_{-50}$ & $140^{+67}_{-50}$ & $217^{+83}_{-74}$  & $1.08^{+0.46}_{-0.41}$  & $53^{+27}_{-29}$ & $>0.63$/$>1.3$ & $1.0/88$\\
	\hline
IPs$^c$ & $115\pm9$ & $107\pm16$ & $80\pm7$  & $0.71\pm0.04$ & - & -&-\\
non-mCVs$^d$ & $62\pm18$ & $438\pm85$ & $95\pm19$  & $0.27\pm0.06$  & - & -&-\\

\hline
  \end{tabular}

\end{center}
  \scriptsize{a \& b: The $T_{\rm max}$ and the $M_{\rm WD}$ derived from the $I_{7.0/6.7}$ values using the $I_{7.0/6.7}$--$T_{\rm max}$--$M_{\rm WD}$ relations by \citet{xu2019}, see text and Figure 3 for details.  \\ c \& d: averaged value of IPs and non-mCVs in the Solar vicinity \citep{xu2016}. }
\end{table}
\endgroup

\section{Iron Line Diagnostics for CV Populations}

\subsection{Methodology}

The maximum temperature ($T_{\rm max}$) and the mass of WD ($M_{\rm WD}$) in CVs can be related via $T_{\max} =\frac{3}{8}\frac{\mu m_H}{k}\frac{GM_{\rm WD}}{R_{\rm WD}}$ for IPs  \citep[e.g.,][]{frank2002}, and  $T_{\max}=\alpha\frac{3}{16}\frac{\mu m_H}{k}\frac{GM}{R}$, where $\alpha=0.65\pm0.07$ for non-mCVs  \citep{yu2018}. However, \textit{Chandra} spectra alone are not robust for the measurement of $T_{\rm max}$ owing to the limited sensitivity of {\it Chandra} at energies above $\sim8$ keV. 
Fortunately, the line ratio $I_{7.0}/I_{6.7}$ can be taken as a sensitive diagnostic for $T_{\rm max}$. The $I_{\rm 7.0}/I_{\rm 6.7}$--$T_{\rm max}$ and $I_{\rm 7.0}/I_{\rm 6.7}$--$M_{\rm WD}$ relations have been built and examined in detail for 25 non-mCVs and IPs in the Solar vicinity based on \textit{Suzaku} and \textit{NuSTAR} observations (e.g., \citealt{xu2016,yu2018,xu2019}). Furthermore, the relation of non-mCVs has been applied to the Galactic bulge X-ray emission (GBXE) to constrain the mean WD mass in CVs \citep[$\sim0.8~M_\odot$;][]{yu2018}.

\begin{figure}[htbp]
  \centering
 \includegraphics[scale=0.3]{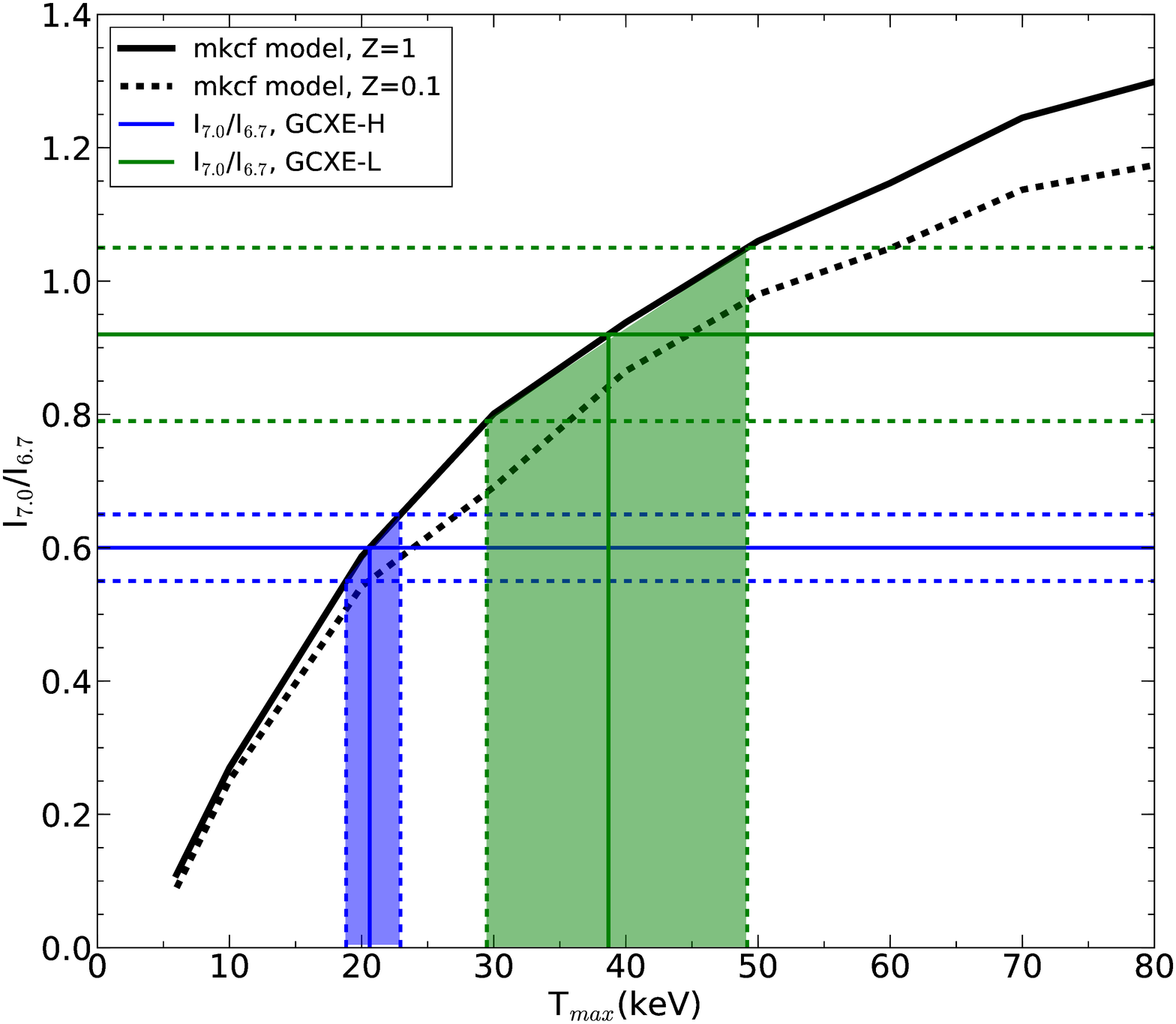}
 \includegraphics[scale=0.3]{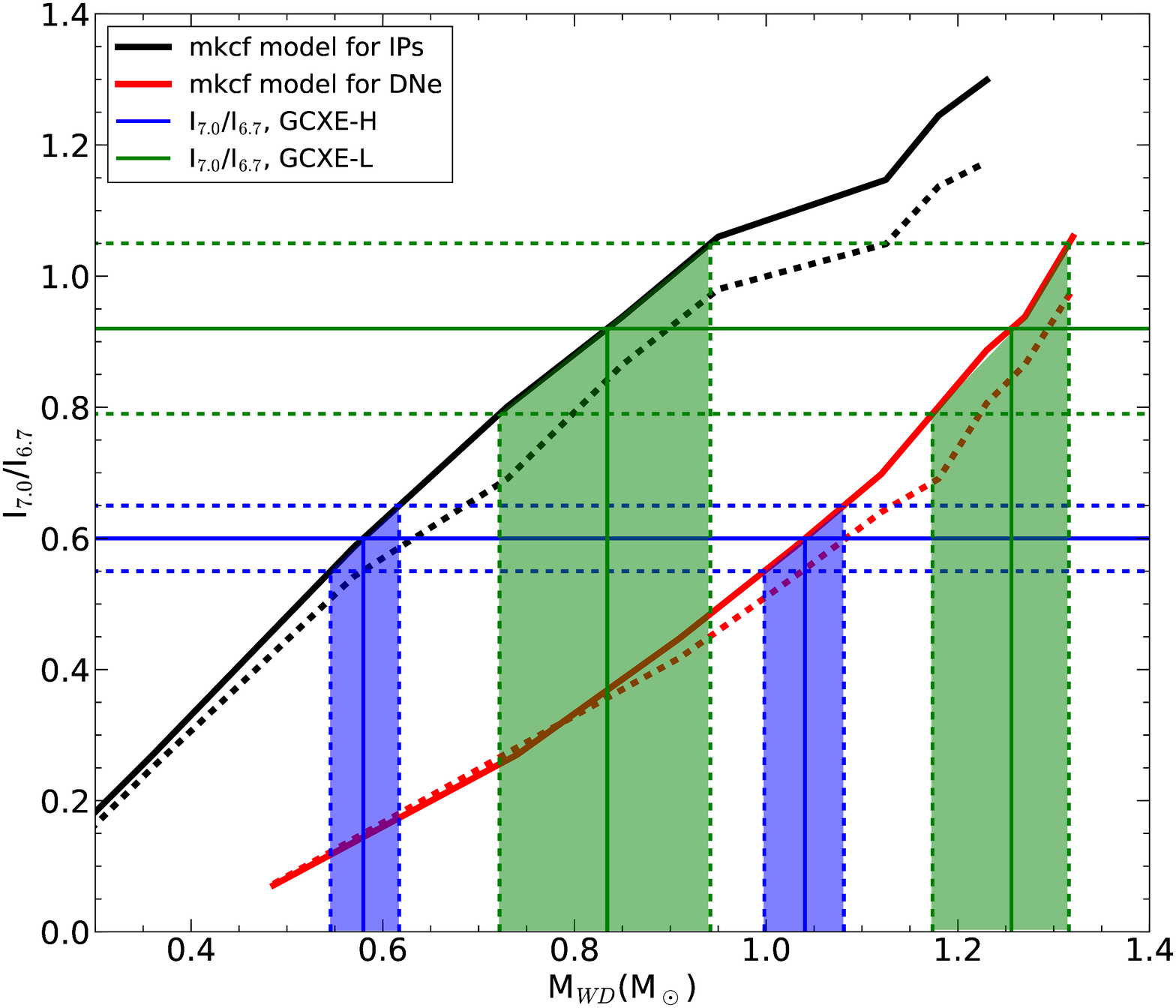}
  \caption{\textit{Upper Panel}: The $I_{\rm 7.0}/I_{\rm 6.7}$--$T_{\rm max}$ relation. The solid and dashed black curves are the predicted relations by mkcflow models of different metallicities (Z = 1 and 0.1 solar value, respectively). The green (blue) horizontal solid and dotted lines mark the measured $I_{\rm 7.0}/I_{\rm 6.7}$ and 90\% uncertainty ranges of GCXE-L (GCXE-H) sources. The green (blue) vertical strips indicate the respective $T_{\rm max}$ values. \textit{Lower Panel}: The $I_{\rm 7.0}/I_{\rm 6.7}$--$M_{\rm WD}$ relations. The black (red) curves are the predicted relations by mkcflow for IPs (non-mCVs). Other lines are as described in the insert.}
\label{fig:ratio}
\end{figure}

To provide a useful diagnostic for the CV populations in the NSC, we incorporate the most recent $I_{\rm 7.0}/I_{\rm 6.7}$--$T_{\rm max}$ and $I_{\rm 7.0}/I_{\rm 6.7}$--$M_{\rm WD}$ relations by \citet{xu2019}, and plot them in Figure~\ref{fig:ratio} as solid and dashed curves for $Z = 1$ and $Z=0.1$ solar values, respectively.
To constrain the metallicity of NSC CVs, we simulate a series of CV spectra using the rmf and arf files of the observed \textit{Chandra} spectrum with metallicities ranging from 0.1 to 2 solar value, assuming as input the mkcflow model (which is generally used to fit the CV spectra, see e.g., \citealt{mukai2017}) with $T_{\rm max}$=40 keV, and 3--8 keV flux and the exposure time same as in the observed spectrum. We find that to reproduce the observed EWs of the GCXE spectrum, the simulated spectrum based on the mkcflow model requires a metallicity $Z > 0.6$ at 90\% confidence level. 
This is consistent with the expectation that the GC stellar populations are predominately of a solar or even super-solar metallicity.

\subsection{$M_{\rm WD}$ of CVs in the NSC}

As shown in Figure 2, the cumulative spectra of GCXE and CHXE sources show significant Fe lines, which represent the average line strengths of the constituent sources, presumably CVs. 
What's more, the luminosity function of the detected NSC sources can be described by $N(>L)\propto L^{-1.63^{+0.16}_{-0.15}}$ \citep{zhu2018}. Such a steep luminosity function implies that the cumulative spectra represent the average properties of the majority of point sources, i.e., the less luminous CVs in the NSC. 
Therefore, we can employ the $I_{\rm 7.0}/I_{\rm 6.7}$--$T_{\rm max}$--$M_{\rm WD}$ relations in Section 3.1 to infer the characteristic shock temperature and mean WD mass of CVs in the NSC.   
         
We take the relation with $Z=1$ as fiducial\footnote{Note the GCXE should have a $Z > 0.6$, as discussed in the last section}, noting that the uncertainty associated with any reasonable range of metallicity in the NSC should be small compared to the statistical errors in the measured $I_{\rm 7.0}/I_{\rm 6.7}$ values.

From Table 1, the GCXE (i.e., the H+L) sources have $I_{\rm 7.0}/I_{\rm 6.7}$ of $0.71\pm0.06$, which is consistent with $I_{\rm 7.0}/I_{\rm 6.7}$ = $0.65\pm0.20$ of the CHXE sources. 
Both values are comparable to the mean $I_{\rm 7.0}/I_{\rm 6.7}$ of IPs in the Solar vicinity ($0.71\pm0.04$, see Table 1 and \citealt{xu2016}), but are significantly higher than the mean of non-mCVs in the Solar vicinity ($0.27\pm0.06$). 
Using the $I_{\rm 7.0}/I_{\rm 6.7}$--$T_{\rm max}$ relation in Figure 3, the average $T_{\rm max}$ of H, L, GCXE and CHXE sources can be estimated as $21_{-3}^{+3}$ keV, $38^{+11}_{-9}$keV, $26_{-3}^{+2}$ keV and $23_{-8}^{+10}$ keV, respectively. Notably, $T_{\rm max}$ of the L sources is comparable to the shock temperature of $43_{-9}^{+11}$ keV measured by {\it NuSTAR} (derived from the $M_{\rm WD}$ measurements with IPM model by \citealt{hailey2016}); $T_{\rm max}$ of the H, GCXE and CHXE sources are consistent with each other, but lower than the \textit{NuSTAR} measurement.
 
According to the $I_{\rm 7.0}/I_{\rm 6.7}$--$M_{\rm WD}$ relation in Figure 3, the average $M_{\rm WD}$ of the H, L, H+L and CHXE sources can be constrained as $0.57^{+0.05}_{-0.02}M_{\odot}$, $0.83^{+0.11}_{-0.11}M_{\odot}$, $0.65_{-0.02}^{+0.07}M_{\odot}$ and $0.61_{-0.14}^{+0.19}M_{\odot}$ if they were mostly IPs, and $1.04^{+0.04}_{-0.04}M_{\odot}$, $1.25^{+0.07}_{-0.08}M_{\odot}$, $1.11_{-0.03}^{+0.06}M_{\odot}$ and $1.07_{-0.16}^{+0.13}M_{\odot}$ if they were predominately non-mCVs. 

\begin{figure}[htbp]
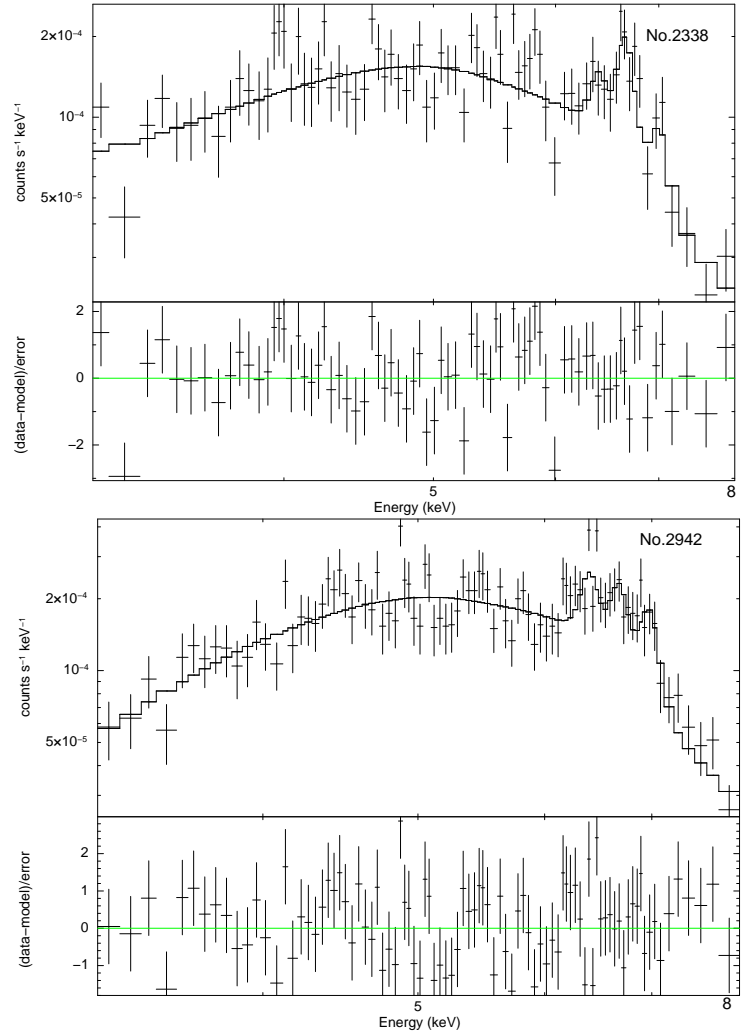

  \centering
 \includegraphics[scale=0.4,angle=270]{src2}
 \includegraphics[scale=0.4,angle=270]{src3}
  \caption{ The cumulative \textit{Chandra} ACIS-S spectra of two brightest sources in the GCXE region. {\it Upper panel}: No.2338. {\it Lower panel}: No.2942. The spectra are fitted by an absorbed bremsstrahlung plus a 3-Gaussian model to characterize the three Fe emission lines. The data error bars are of 1-$\sigma$.}
  \label{fig:specindiv}
\end{figure}

In the above analysis, we assume that the mean line ratio and its error can represent the typical range of the individual sources. To verify this assumption, we further inspect the spectra of the brightest individual sources. We extract and fit the cumulative spectra of two brightest sources (Source No.2338 and No.2942 in the NSC X-ray source catalog of \citealt{zhu2018}, each having more than 1800 net counts in the combined ACIS-S data.) in the GCXE region with the same procedure. Both sources belong to the H group, and their J2000 coordinates are R.A.=17:45:41.498, Decl.=-28:58:14.83 for No.2338 and R.A.=17:45:48.948, Decl.=-28:57:51.73 for No.2942, respectively. The spectra and fitting results are presented in Figure 4 and Table 1. It can be seen that the spectra of No.2338 and No.2942 both show significant Fe lines and bremsstrahlung--like continuum, which are typical for CVs.  The $I_{\rm 7.0}/I_{\rm 6.7}$ of No.2338 ($0.30^{+0.28}_{-0.17}$) is consistent with that of non-mCVs, and the $I_{\rm 7.0}/I_{\rm 6.7}$ of No.2942 ($1.08^{+0.46}_{-0.41}$) is consistent with that of IPs in the Solar vicinity. We further examine the 2-8 keV light curve of the two sources, following the procedure described in \citet{zhu2019}. The results indicate that the flux of No.2338 shows a variability up to an order of magnitude ($10^{-6}$ to $10^{-5}$ photon cm$^{-2}$ s$^{-1}$), which are comparable to non-mCVs, especially dwarf novae in the Solar vicinity (e.g., \citealt{wada2017}). 
It is noteworthy that the spectral shapes appear slightly different for the spectra extracted in the high-flux and low-flux states, however no significant deviation of the fitted $I_{\rm 7.0}/I_{\rm 6.7}$ can be measured due to relatively large statistical uncertainties. On the other hand, the flux of No.2942 remains constant ($\sim7\times10^{-6}$ photon cm$^{-2}$ s$^{-1}$) over the last 19 years, just like IPs in the Solar vicinity. 
Given the high(low) variabilities and the low(high) $I_{\rm 7.0}/I_{\rm 6.7}$ value, No.2338(No.2942) is likely a non-mCVs (an IP).
The WD masses can be derived to be $0.36^{+0.22}_{-0.11}M_{\odot}$ ($0.77^{+0.25}_{-0.30}M_{\odot}$) for No.2338 if it is an IP (non-mCV) ; and $>0.63M_{\odot}$ ($>1.3M_{\odot}$) for No.2942 if it is an IP (non-mCV).\footnote{The upper limit of WD mass of No.2942 is not constrained because the upper limit of $I_{\rm 7.0}/I_{\rm 6.7}$ is beyond the range of the $I_{\rm 7.0}/I_{\rm 6.7}$--$M_{\rm WD}$ relation, see Figure 3 for details.}.

The implications of these values are addressed below.

\section{Discussion}

\subsection{The Nature and Formation Channel of NSC CVs}
The $I_{\rm 7.0}/I_{\rm 6.7}$ and the derived $T_{\rm max}$ and $M_{\rm WD}$ values provide important clues on the nature of NSC CVs, as discussed below.
Given the 2-10 keV luminosity range, H sources are comparable to IPs in Solar vicinity. What's more, the $I_{\rm 7.0}/I_{\rm 6.7}$ value ($0.60\pm0.05$) is a little bit lower than that of IPs in the Solar vicinity ($0.71\pm0.04$, see \citealt{xu2016}). The most natural nature of H sources are then a mixture of IPs and non-mCVs, as suggested by the above analysis on No.2338 and No.2942.

The situation for L sources is a little bit different. Their luminosities are consistent with non-mCVs in the Solar vicinity  \citep[e.g.,][]{rev09,xu2016,yu2018,byckling2010,reis2013}, but their mean $I_{\rm 7.0}/I_{\rm 6.7}$ ($0.92\pm0.22$) is significantly higher than those of the Solar vicinity non-mCVs ($I_{\rm 7.0}/I_{\rm 6.7}$$\sim0.2$, see \citealt{xu2016}). 
The inferred $M_{\rm WD}$ of the L sources, in particular, is to be contrasted with the average $M_{\rm WD}$ $\approx 0.8M_{\odot}$ of the Galactic bulge CVs with similar X-ray luminosities (which were suggested to be mostly non-mCVs with averaged $I_{\rm 7.0}/I_{\rm 6.7}$ $\sim0.3$, see \citealt{yu2018}). 
Now there are two possibilities, either L sources are non-mCVs with mean $M_{\rm WD}$$>1.0M_{\odot}$, or they are low luminosity IPs. The formation of these CVs could be related to the dynamical encounters in the NSC, which are briefly discussed as follows.

Theoretically, the most important dynamical effects includes the gravitational influence of the SMBH, the mass segregation and close encounters between stars. The mass segregation tends to bring massive stars and binaries to the vicinity of the SMBH; the close stellar encounters can selectively bring massive stars into binaries, and alter the orbits of binaries  \citep[e.g.,][]{heggie1975,hills1975,hut1993}. As a result, the WDs of the descendent CVs are supposed to be significantly higher from their field counterparts, i.e., binaries subject only to secular stellar evolution. 
This scenario is also supported by the high $EW$ of the NSC CVs, which indicates that the donor star must be relatively metal-rich ($Z>0.6$), which is at odds with the typical metal-poor stellar populations in globular clusters, if the CVs were originally formed in globular clusters and sequently fallen into the NSC \citep[e.g.,][]{tremaine1975,antonini2012}. 
This strongly suggests that the CVs presently detected in the NSC have been reprocessed (their companions have been exchanged) after their infall, which is supported by numerical simulations \citep{panamarev2018}.
On the other hand, this scenario may also favor the formation of IPs. The dynamical exchange in globular clusters and in the NSC could shrink the orbital separation, and enhance the population of close binaries. Such an effect would naturally lead to the formation of tighter post-common envelope binaries, which were suggested to favor the formation of mWDs \citep{briggs2018}.

\subsection{Contribution of Resolved CVs to the 20-40 keV CHXE}

The unresolved 20--40 keV CHXE detected by {\it NuSTAR} was suggested to predominantly arise from a large number of IPs \citep{perez2015,hailey2016}.  
The contribution of {\it Chandra}-resolved CVs to the CHXE can be estimated by extrapolating the \textit{Chandra} spectra to 40 keV, assuming a cooling flow model with $T_{\rm max}$$=23_{-8}^{+10}$ keV, which is obtained from the $I_{\rm 7.0}/I_{\rm 6.7}$--$T_{\rm max}$ relation for $I_{\rm 7.0}/I_{\rm 6.7}$$\approx0.65\pm0.20$ (Table 1).
The thus derived 20--40 keV flux is $(1.7\pm0.4)\times10^{-13}$ erg~cm$^{-2}$~s$^{-1}$, which is $(24\pm6)$\% of the total flux measured by \textit{NuSTAR} ($\sim 7\times10^{-13}$ erg cm$^{-2}$~s$^{-1}$, see \citealt{hailey2016}).
A more delicate estimate comes from separately extrapolating the spectra of the GCXE-H and GCXE-L sources (the normalization of the spectra are rescaled according to the stellar mass enclosed in the GCXE and the CHXE regions), for them having different $I_{\rm 7.0}/I_{\rm 6.7}$ (hence different $T_{\rm max}$).  
This results in a 20--40 keV flux of $(2.4\pm0.5)\times10^{-13}$ erg cm$^{-2}$~s$^{-1}$, with $\sim$$60\%$ from H sources and $\sim$$40\%$ from L sources.
The two estimates are consistent with each other, and only account for $\lesssim 42\%$ of the \textit{NuSTAR} flux. This deficit might be explained if there is a large population of less luminous CVs, and/or milli-second pulsars which were proposed to be abundant in the Galactic center region \citep[e.g.,][]{eckner2018}.

\section{Summary}
We have investigated the combined X-ray spectra of the CVs located in the Galactic center region based on archival \textit{Chandra} ACIS-S observations to trace their mean WD masses. 
We focus on the nuclear star cluster (NSC) region, more specifically, the half-circular region with a projected galacto-centric radius $R = 250\arcsec$ and Galactic latitude $b>0$, defined as the Galactic center X-ray emission (GCXE) region. We divide the point sources detected therein into GCXE-H (with $L_{\rm 2-10keV} > 6\times10^{31}{\rm~erg~s^{-1}}$) and GCXE-L (with $L_{\rm 2-10keV} < 6\times10^{31}{\rm~erg~s^{-1}}$) subgroups according to their X-ray luminosities. We also examine the Chandra sources falling within the central hard X-ray emission south-west (CHXE-SW) region as defined in \citet{perez2015}.
Our main results and conclusions are as follows.\\
a) The CVs with $L_{\rm 2-10 keV}>6\times10^{31}$ erg s$^{-1}$ ($L_{\rm 2-10 keV}\sim1-6\times10^{31}$ erg s$^{-1}$) in the NSC have a mean $T_{\rm max}$ of $21_{-3}^{+3}$ ($38^{+11}_{-9}$) keV, which corresponds to a mean WD mass of $0.57^{+0.05}_{-0.02}M_{\odot}$ ($0.83^{+0.11}_{-0.11}M_{\odot}$) if the dominate CV population are IPs, or $1.04^{+0.04}_{-0.04}M_{\odot}$ ($1.25^{+0.07}_{-0.08}M_{\odot}$), if they are non-mCVs, respectively;\\
b) The \textit{Chandra} detected point sources can contribute $\lesssim42\%$ of the 20-40 keV CHXE;\\
c) The massive WDs in the CVs likely result from dynamical exchanges in the NSC.

\acknowledgements
The authors thank the anonymous referee for constructive comments that helped improve this paper. This work is supported by National Science Foundation of China through grants 11873029, 11473010, 11303015, 11773015, 11133001, and the National Key Research and Development Program of China (2016YFA0400803 and 2017YFA0402703). We acknowledge useful comments from Shuo Zhang on the CHXE. This work made use of data from the NuSTAR mission, a project led by the California Institute of Technology, managed by the Jet Propulsion Laboratory, and funded by the National Aeronautics and Space Administration.

\facility{CXO (ACIS), NuSTAR}.

\end{document}